\documentclass[a4paper,english,showpacs,superscriptaddress,aps,amsmath,amssymb,twocolumn,prl]{revtex4}
\usepackage[T1]{fontenc}
\usepackage[latin9]{inputenc}
\usepackage{amsmath}
\usepackage{graphicx}
\usepackage{amssymb}

\makeatletter

\providecommand{\tabularnewline}{\\}

\@ifundefined{textcolor}{}
{%
 \definecolor{BLACK}{gray}{0}
 \definecolor{WHITE}{gray}{1}
 \definecolor{RED}{rgb}{1,0,0}
 \definecolor{GREEN}{rgb}{0,1,0}
 \definecolor{BLUE}{rgb}{0,0,1}
 \definecolor{CYAN}{cmyk}{1,0,0,0}
 \definecolor{MAGENTA}{cmyk}{0,1,0,0}
 \definecolor{YELLOW}{cmyk}{0,0,1,0}
 }

\usepackage{bbding}\usepackage{amsfonts}\usepackage{multirow}

\makeatother

\usepackage{babel}

\begin{document}

\title{Gapped Ferromagnetic Graphene Nanoribbons }

\author{D.~Hou}

\affiliation{School of Physics, National Key Laboratory of Crystal Materials,
Shandong University, Jinan 250100, China}

\affiliation{Department of Physics, Renmin University of China, Beijing 100872,
China}

\author{J.~H.~Wei}

\email{wjh@ruc.edu.cn}

\affiliation{Department of Physics, Renmin University of China, Beijing 100872,
China}

\author{S.~J.~Xie}

\email{xsj@sdu.edu.cn}

\affiliation{School of Physics, National Key Laboratory of Crystal Materials,
Shandong University, Jinan 250100, China}

\date{\today}
\begin{abstract}
We theoretically design a graphene-based all-organic ferromagnetic
semiconductor by terminating zigzag graphene nanoribbons (ZGNRs)
with organic magnets. A large spin-split gap with 100\% spin
polarized density of states near the Fermi energy is obtained, which
is of potential application in spin transistors. The interplays
among electron, spin and lattice degrees of freedom are studied
using the first-principles calculations combined with fundamental
model analysis. All of the calculations consistently demonstrate
that although no $d$ electrons existing, the antiferromagnetic
$\pi-\pi$ exchange together with the strong spin-lattice
interactions between organic magnets and ZGNRs make the ground state
ferromagnetic. The fundamental physics makes it possible to
optimally select the organic magnets towards practical applications.
\end{abstract}


\maketitle \emph{Introduction}.---Since the experimental discovery
of graphene in 2004 \cite{Nov04}, graphene-based transistors have
rapidly developed and are considered good candidates for
post-silicon electronics \cite{Sch10}. However, they encounter an
outstanding challenge of how to open a sizable and well-defined band
gap in graphene to control the ON- and OFF-states of the
transistors. One realistic solution is constraining large-area
graphene in one dimension to form graphene nanoribbons (GNRs) which
have band gaps approximately inversely proportional to their widths
\cite{Yang07}. In recent years, this scheme has been experimentally
proved rather practical and thus motivated a new frontier area
called {}"graphene nanoribbon electronics" \cite{Han07,Li08,Chen07}.

Besides the gapped semiconductor properties suitable for
electronics, graphene nanoribbons possess fascinating magnetic
properties for spintronics. Zigzag graphene nanoribbons (ZGNRs) have
spatial spin-polarized ground states with spin moments coupled
ferromagnetically (FM) on the same edge and antiferromagnetically
(AFM) between different edges \cite{Fuj96,Oka01}, which makes them
potential materials for carbon-based spintronic devices avoiding
heavy magnetic atoms. By means of electric field control,
Boron/Nitron doping or edge termination, ZGNRs can exhibit
half-metallicity that may be used for spin injection and filtration
in metallic spintronics\cite{Son06-2,YJL08,Zheng09,sodi}.

On the other side, semiconductor spintronics has been the subject
of many recent studies \cite{Jun06}. The central issue is how to
develop spin transistors with appropriate ferromagnetic semiconductor
materials for non-volatile memory applications \cite{Dat90}. By virtue
of the unique electronic and magnetic properties mentioned above,
GNRs may be ideal materials for spin transistors. However, the difficulty
is that the ferromagnetic state of ZGNRs is not the ground state,
and even worse for transistors, it is a gapless state. Therefore,
the organic integration of electronic and magnetic properties of graphene
for spin transistor is still an open question.

\begin{figure}[htbp]
\includegraphics[width=3.2in]{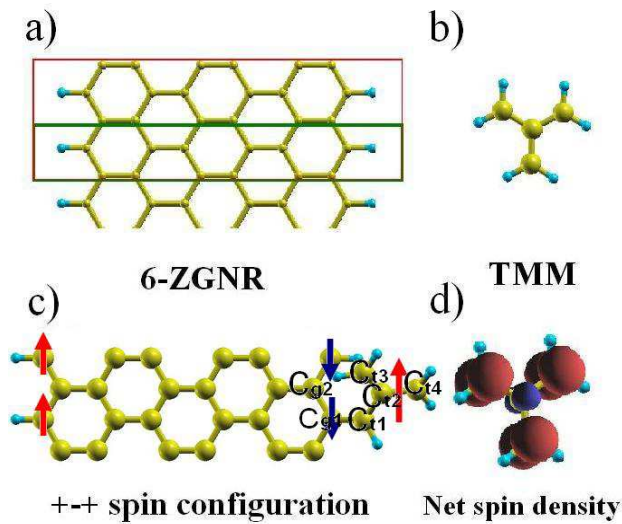}\\
 \caption{(color online) Top view structures of (a) pristine 6-ZGNR, (b) TMM,
(c) TMM terminated 6-ZGNRs, and (d) the net spin density of TMM. The
yellow balls are carbon atoms and the cyanic ones are hydrogen atoms.
The red/blue arrow indicates the up/down spin component.}

\label{fig1}
\end{figure}

In this paper, we design a graphene nanoribbon terminated with
organic magnets to give a possible solution for the above issue. In
order to convert the ground state of ZGNRs from the AFM state to a
gapped FM one, we use edge passivation of graphene nanoribbons by
organic ferromagnetic radicals (one kind of organic magnets, see
Fig.\ref{fig1}). The physics is that the magnetic order of the host
material will be changed when it exchange-couples to the localized
spin of attached organic magnets \cite{Sugawara09}. Comparing with
ZGNRs adsorbing magnetic transition metal atoms on their surfaces in
literatures \cite{Gor08,Rigo09}, these carbon-based materials have
the advantages of much weaker spin-orbit and hyperfine interactions,
which should induce much longer spin coherent length. Furthermore,
owing to the planar structure of the selected organic radical, above
mentioned exchange coupling is tunable by manually rotating the
radical plane to affect the spin polarization near the Fermi energy.
Our results suggest that the controllable graphene-based spin
transistors can be expected.

\emph{Methods}.---Towards the practical applications, we firstly
choose trimethylenemethane (TMM) as the organic magnet, for the
reason that it has been widely investigated and designed for kinds
of possible spintronic applications due to its high magnetic moments
\cite{TMM}. However, as will be demonstrated later, the dependence
of the physics on this specific choice is quite weak. TMM contains 4
carbon and 6 hydrogen atoms with $\pi$-conjugated structure and has
a ground state of triplet biradical, as shown in Fig.\ref{fig1}b.
Two unpaired electrons in TMM provide 2$\mu_{B}$ spin moments in
total distributing via molecular orbital over all the carbon atoms.
At its spin triplet ground states, three outer carbon atoms have the
same kind of net spin density, while the inner one has the opposite
(see Fig.\ref{fig1}d). When attached to one of the ZGNRs' edges
already possessing spin moments, TMM will strongly modify the
properties of ZGNRs.

6-ZGNR terminated by hydrogen atoms is used as pristine ZGNR as
shown in Fig.\ref{fig1}a. The unit cell of 6-ZGNR contains 12 carbon
atoms and 2 hydrogen atoms (see the figure inside the green
rectangle in Fig.\ref{fig1}a). The length of periodic direction of
the unit cell is 2.46 $\mathrm{\textrm{\AA}}$ (experiment value),
while the repeated images of the molecular are separated by vacuum
more than 8 $\mathrm{\textrm{\AA}}$ thick along the other two
directions. To avoid direct interaction between images of terminated
TMM, the zigzag direction of the unit cell is doubled (see the
region inside the red rectangle in Fig.\ref{fig1}a). In this
supercell, TMM replaces one of the four hydrogen atoms, making the
system terminated by TMM and hydrogen alternately on one edge (e.g.
the right edge). The top view of one possible relaxed structure
(that will be proved the ground state) is shown in Fig.\ref{fig1}c.
Please be noted that the TMM biradical is not coplanar with ZGNR
after atomic optimization.

The first-principles calculations are performed using Density
Functional Theory (DFT) method implemented with \texttt{SIESTA}
code\cite{Siesta} using the Perdew-Burke-Ernzerhof
exchange-correlation functional\cite{PBE}. Double-$\zeta$ plus
polarization function (DZP) basis set is used, together with a mesh
cutoff of 200 Ry and norm-conserving pseudopotentials. All the atoms
in the supercell are fully relaxed to fulfill the energy and force
convergence of 10$^{-5}$ eV and 0.01 eV/$\mathrm{\textrm{\AA}}$
respectively. The optimized structure parameters with different
initial spin configurations are crosschecked using a plane-wave
method implemented in \texttt{Quantum Espresso} code\cite{QE}, with
the same convergence criteria and a plane wave kinetic energy cutoff
of 30 Ry. The identical results ensure the validity of the
simulation parameters used, and other data are handled using
\texttt{SIESTA} code solely.

\emph{Results.}---In what follows, we use {}"\texttt{++}" and
{}"\texttt{+-}" to respectively denote the FM and AFM spin
configurations, where the first symbol denotes the left edge spin
moment while the second denotes the right one of the graphene
ribbon. Without losing generality, we designate {}"\texttt{+}/-" to
represent up/down spin in what follows. The convergence of the
simulation parameters is firstly checked by electronic property
calculations of a pristine 6-ZGNR with a 1$\times$1$\times$2
supercell. The {}"\texttt{+-}" order is confirmed to be the ground
state with a net spin moments of 0.26 $\mu_{B}$ localized on each
edge atom, and its total energy is 26 meV/supercell lower than the
{}"\texttt{++}" one. These values are consistent with previously
reported results and used as reference data \cite{Fuj96,Oka01}.

The TMM units are then attached to the right side of the ZGNR. Since
the spin moments on TMM and ZGNRs can align either parallel or
anti-parallel, the {}"\texttt{++}" and {}"\texttt{+-}" spin
configurations will extend to four possible ones: {}"\texttt{+-+}",
{}"\texttt{++-}",{}"\texttt{+-{}-}" and {}"\texttt{+++}", with the
rightmost symbol representing the net spin moments on TMM. Full
atomic relaxation calculations are done separately using these four
possible spin configurations to find the ground state. It is found
that both the bond length and the dihedral angle between the ZGNR
and TMM are varied with different spin configurations ( we will
discuss this point in details later). The relative total energy (
taking $E_{\texttt{+-+}}$ as reference), converged net spin moments
and optimized bond structures are summarized in Tab.\ref{tab1}.
Crosschecking with \texttt{Quantum Espresso} presents highly
consistent results, as shown in Tab.\ref{tab1}.

\begin{table*}[tp] \caption{Relative
total energy, converged net spin moments on selected carbon atoms,
optimized bond lengths and spin angles for all possible spin
configurations of TMM terminated 6ZGNR. The meanings of the symbols
for the carbon atoms are depicted in Fig.\ref{fig1}c. The values in
brackets are corresponding results obtained by \texttt{Quantum
Espresso} code.}

\label{tab1}\centering \begin{tabular}{|c|c|c|c|c|c|c|c|c|c|c|c|c|}
\hline \multirow{2}{*}{Label}  &
\multirow{2}{*}{E-E$_{\texttt{+-+}}$ (meV)} &
\multicolumn{6}{|c|}{Net Spin Moments ($\mu_{B}$)} &
\multicolumn{4}{|c|}{Bond Lengths ($\mathrm{\textrm{\AA}}$)}& Spin
Angle\tabularnewline \cline{3-13} &  & C$_{g1}$  & C$_{g2}$ &
C$_{t1}$  & C$_{t2}$  & C$_{t3}$  & C$_{t4}$  & C$_{g1}$-C$_{t1}$ &
C$_{t1}$-C$_{t2}$  & C$_{t2}$-C$_{t3}$  & C$_{t2}$-C$_{t4}$ &
C$_{g1}$-C$_{t1}$\tabularnewline \hline \texttt{+-+} & 0  & -0.08  &
0.09  & 0.25  & -0.11 & 0.60  & 0.51  & 1.39(1.38)  & 1.48(1.48)  &
1.42(1.41)  & 1.40(1.40) & 180\textordmasculine \tabularnewline
\hline \texttt{++-} & 36 & 0.08  & -0.08  & -0.24 & 0.11  & -0.60  &
-0.51 & 1.39(1.38) & 1.48(1.48)  & 1.42(1.41) & 1.40(1.40) &
180\textordmasculine \tabularnewline \hline \texttt{+-{}-} & 578 &
-0.24  & 0.02  & -0.63 & 0.11  & -0.68  & -0.68  & 1.48(1.48)  &
1.43(1.43) & 1.42(1.42)  & 1.42(1.42) & 70\textordmasculine
\tabularnewline \hline \texttt{+++} & 612 & 0.23 & -0.02 & 0.63  &
-0.11  & 0.68  & 0.68  & 1.48(1.48) & 1.43(1.43) & 1.42(1.42)  &
1.42(1.42) & 70\textordmasculine \tabularnewline \hline
\end{tabular}
\end{table*}

As presented in Tab.\ref{tab1}, the "\texttt{+-+}" state is the
ground state. It has a total energy of 36 meV lower than that of
"\texttt{++-}" state, 578 meV than that of "\texttt{+-{}-}" state
and 612 meV than that of "\texttt{+++}" state, respectively. These
data indicate that, in the ground state of the radical terminated
graphene nanoribbon, the net spin moments of TMM radical
antiferromagnetically couples to the nearest edge spin of ZGNR. As
the energy difference between pristine "\texttt{+-}"{}"\texttt{++}"
ZGNR is 26 meV, the present coupling is much stronger than the
interedge superexchange of ZGNR\cite{Jun09}. A relative result was
reported just recently by Atodiresei et al. focusing on the spin
polarization of nonmagnetic benzene, cyclopentadienyl radical and
cyclooctatetraene molecule adsorbed onto a ferromagnetic 2 ML
Fe/W(110) surface. It is found that at the organic molecule site an
inversion of the spin polarization occurs with respect to the
ferromagnetic surface, resulting from the antiferromagnetic coupling
between $\pi$-electrons of the molecule and the $d$-electrons of Fe
atoms\cite{Ato}. While in the present case, the strong
antiferromagnetic coupling presented here results from the
$\pi$-$\pi$ interactions.

Comparing the corresponding structures of the four spin
configurations manifests the close connection between the spin
coupling and the variation of the ZGNR-TMM bond length (or dihedral
angle). It can be seen from Tab.\ref{tab1} that the changes of bond
lengths and spin moments among the four possible spin configurations
mainly take place around the C$_{g1}$-C$_{t1}$ bond. When TMM and
ZGNR are coupled antiferromagnetically ("\texttt{+-+}" and
"\texttt{++-}" states), the length of the C$_{g1}$-C$_{t1}$ bond is
shortened about 0.1$\mathrm{\textrm{\AA}}$ relative to those of
"\texttt{+-{}-}" and "\texttt{+++}" states. It is reasonable to
attribute such a large distortion to the softness of the all-organic
system and the strong spin-lattice interaction. It is further found
that, if the dihedral angle is disturbed, the bond length will
changes correspondingly. As the $\pi$ electron orbital trends to be
perpendicular to the $\sigma$ bonds surface, the dihedral angle
reflects the spin angle of $\pi$-$\pi$ electrons on C$_{g1}$ and
C$_{t1}$. Therefore, the calculation reveals that the $\pi$-$\pi$
spin coupling depends not only on the bond length $d$ but also on
the spin angle $\theta$. We thus propose a spin-lattice
(spin-phonon) coupling model to illustrate the physics as follows,
\begin{equation}
\begin{split}H_{\mathrm{T}} & =H_{c}+J_{0}(1-\alpha u)\vec{S}_{g1}\cdot\vec{S}_{t1}+\frac{k}{2}u^{2}\\
 & \end{split}
\label{eq:sp-lat}\end{equation} where $H_{c}$ denotes the invariant
part (except the spin-spin coupling between C$_{g1}$ and C$_{t1}$) before
and after spin-lattice interaction involved. The second term is the
$\pi$-$\pi$ spin interaction modulated by the bond distortion $u$
between C$_{g1}$ and C$_{t1}$, where $J{}_{0}$ is the coupling
constant at $u=0$, $\alpha$ the spin-lattice coupling constant
($\alpha>0$). The third one is the elastic potential energy caused
by the bond distortion. As the magnetism of the terminated radical
is robust, we take a classic approach with
\begin{equation}
\vec{S}_{g1}\cdot\vec{S}_{t1} = S_{g1}S_{t1}\cos\theta
\label{eq:theta}\end{equation}

By taking the differential of the Hamiltonian (\ref{eq:sp-lat}) with
respect to the bond distortion $u$, we can obtain the stable state
under certain fixed angle $\theta$,
\begin{equation}\label{eq:theta-u}
{u}=\frac{\alpha{J_{0}}S_{g1}S_{t1}\cos\theta}{k}=u_0\cos\theta\end{equation}
which indicates that the optimal bond distortion $u$ will change
with the spin angle $\theta$. We then obtain the optimal spin angel
$\theta$ and bond distortion $u$ by minimizing the total energy,
which is given as

\begin{equation} \label{eq:gstate}
\left\{ \begin{aligned}
         \theta &= \pi \\
                  u &= -u_0 <0
                          \end{aligned} \right.
                          \end{equation}

It derives that the ground state is characterized by
antiferromagnetic exchange interaction and a shorter
C$_{g1}$-C$_{t1}$ bond length, which is consistent with our
first-principles calculations as shown in Tab. \ref{tab1}.

\begin{figure}[bpth]
\includegraphics[width=3in]{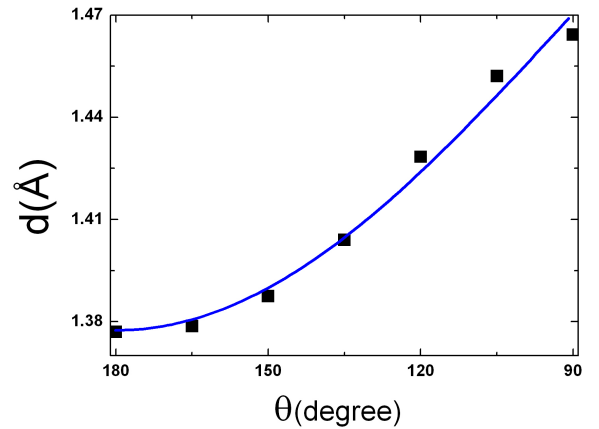}

\caption{(color online) $d-\theta$ relation of the total energy
minimum state for given spin angle $\theta$. The black squares are
calculated using DFT, and the blue line is the fitted line using
equation $d$=$d_{0}+u$=$d_{0}+u$$_{0}$cos$\theta$, where the fitting
parameters $d_{0}$=1.470$\textrm{\AA}$ and
$u_0$=0.093$\textrm{\AA}$.} \label{fig2}
\end{figure}

In order to elucidate the robust physics of the spin-lattice
coupling, we substitute the TMM radicals with CH$_{2}$, the simplest
carbon based organic radical holding the $sp^{2}$ hybridization. In
the first principle calculation, the optimal bond length $d$ between
C$_{g1}$ and C$_{t1}$ for each $\theta$ is obtained by minimizing
the total energy. The calculated $d-\theta$ relation is shown in
Fig. \ref{fig2}, which is well fitted with a cosine type line as
deduced from our model. Therefore, in our proposed all-organic
ferromagnetic graphene structure, the ferromagnetism is provided by
the $\pi-\pi$ spin interaction, while the spin-lattice coupling is
vital for its stability.

\begin{figure}[htbp]
\includegraphics[width=1.6in,height=2in]{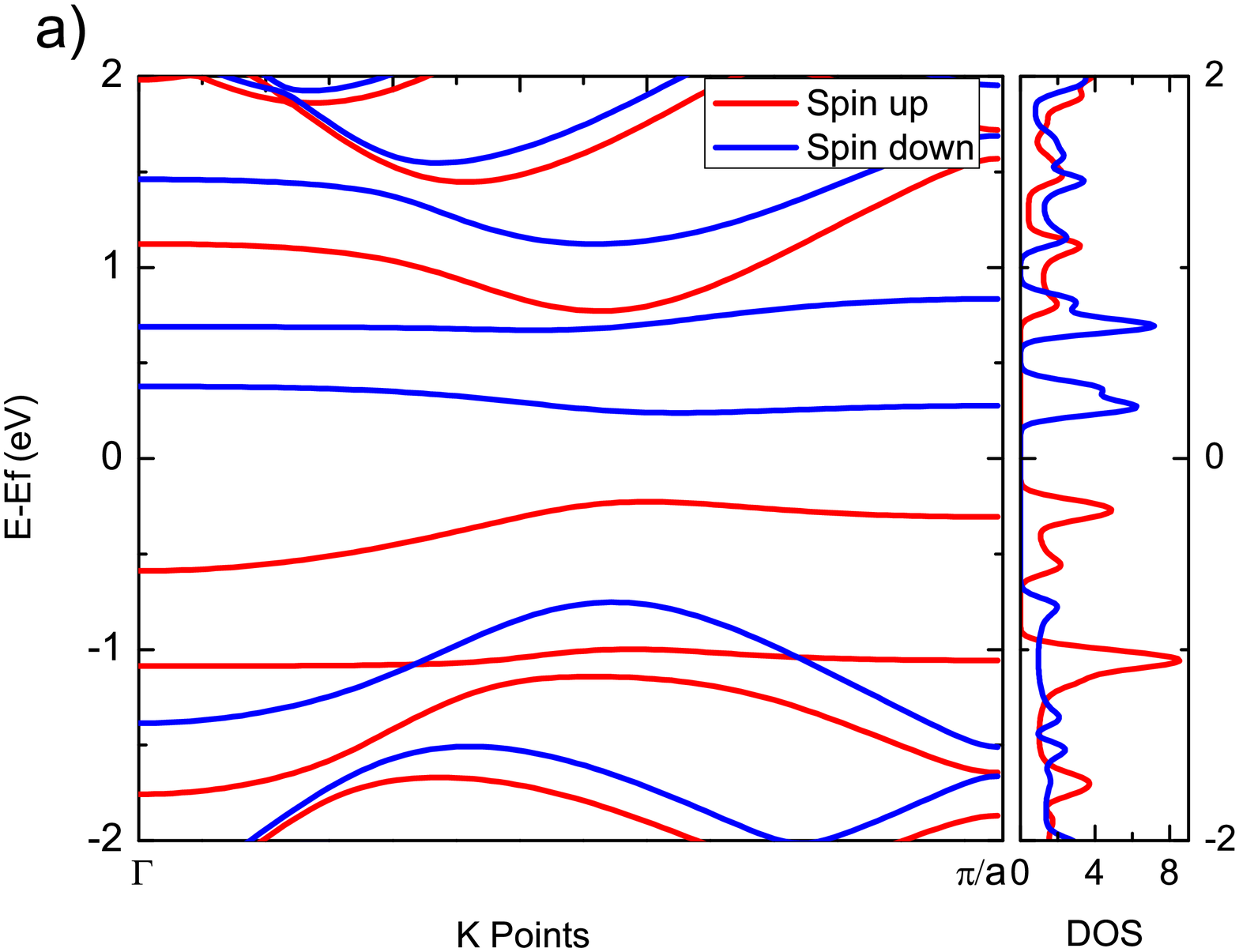}\includegraphics[width=1.6in,height=2in]{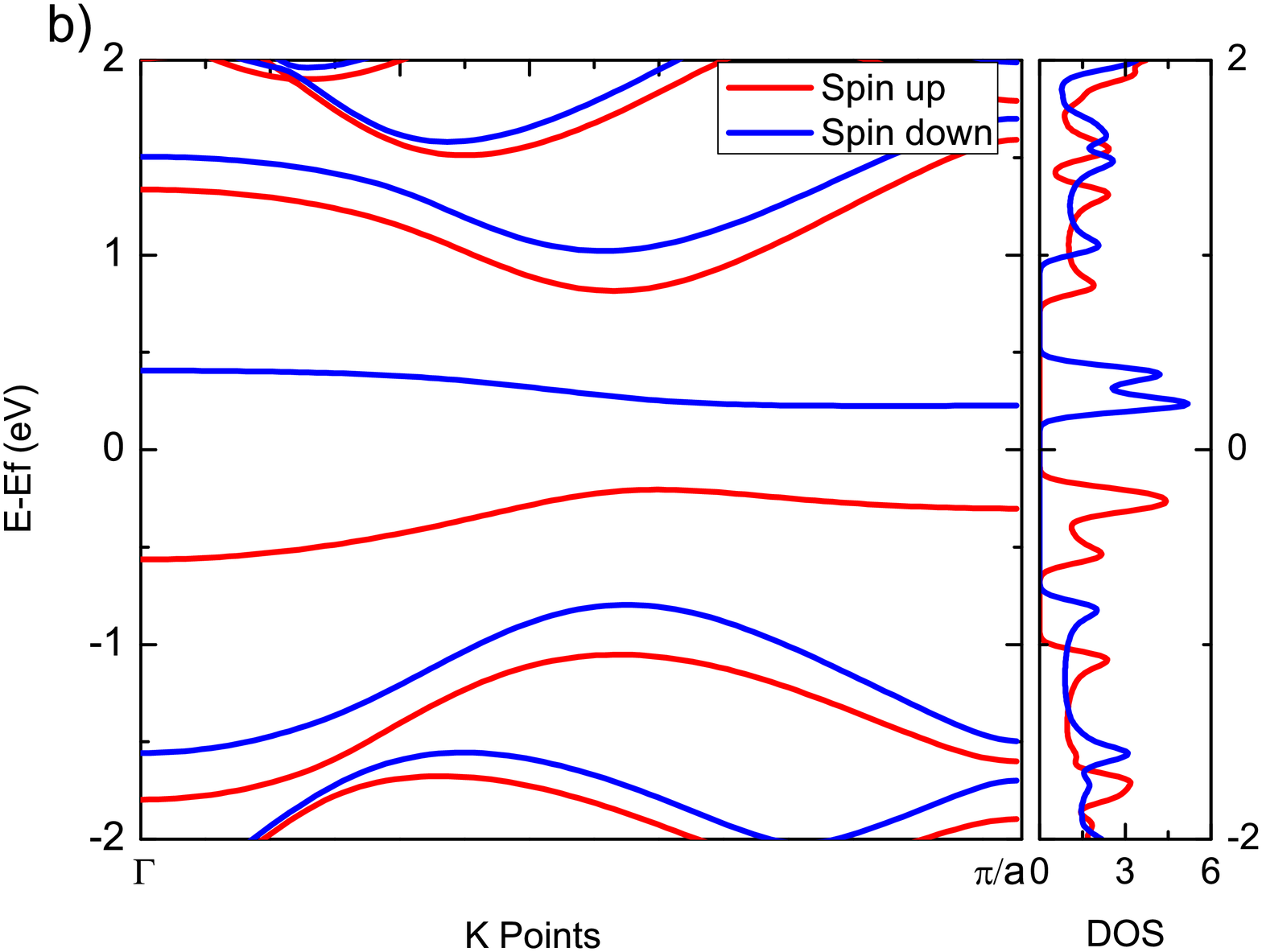}
\includegraphics[width=8cm]{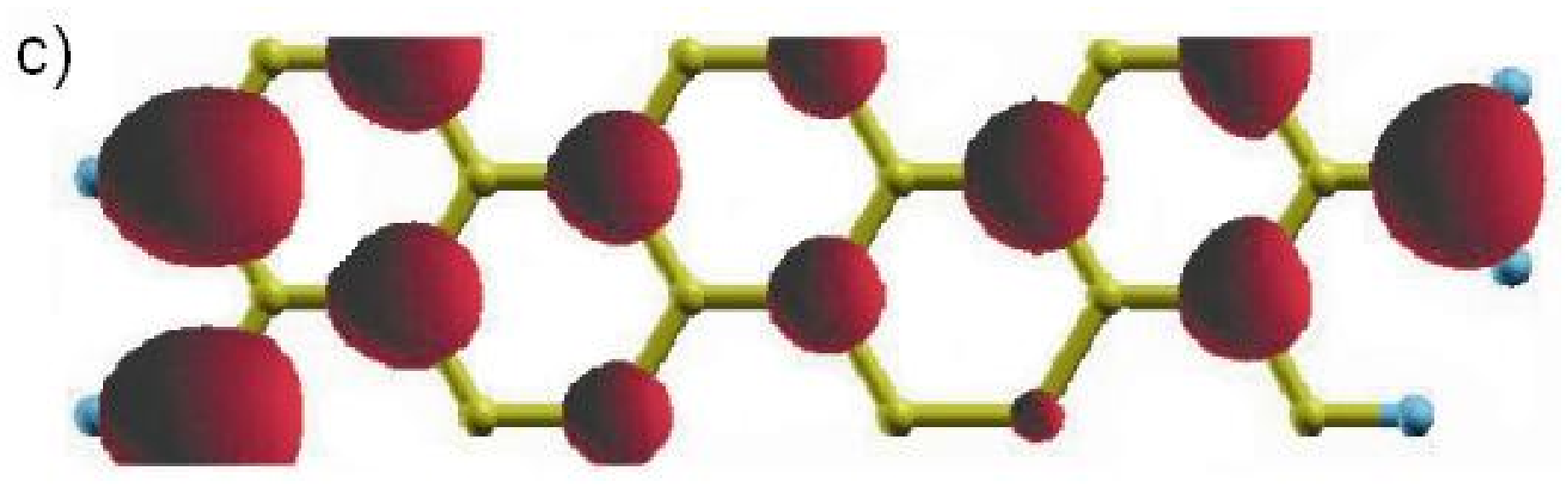}
\includegraphics[width=8cm]{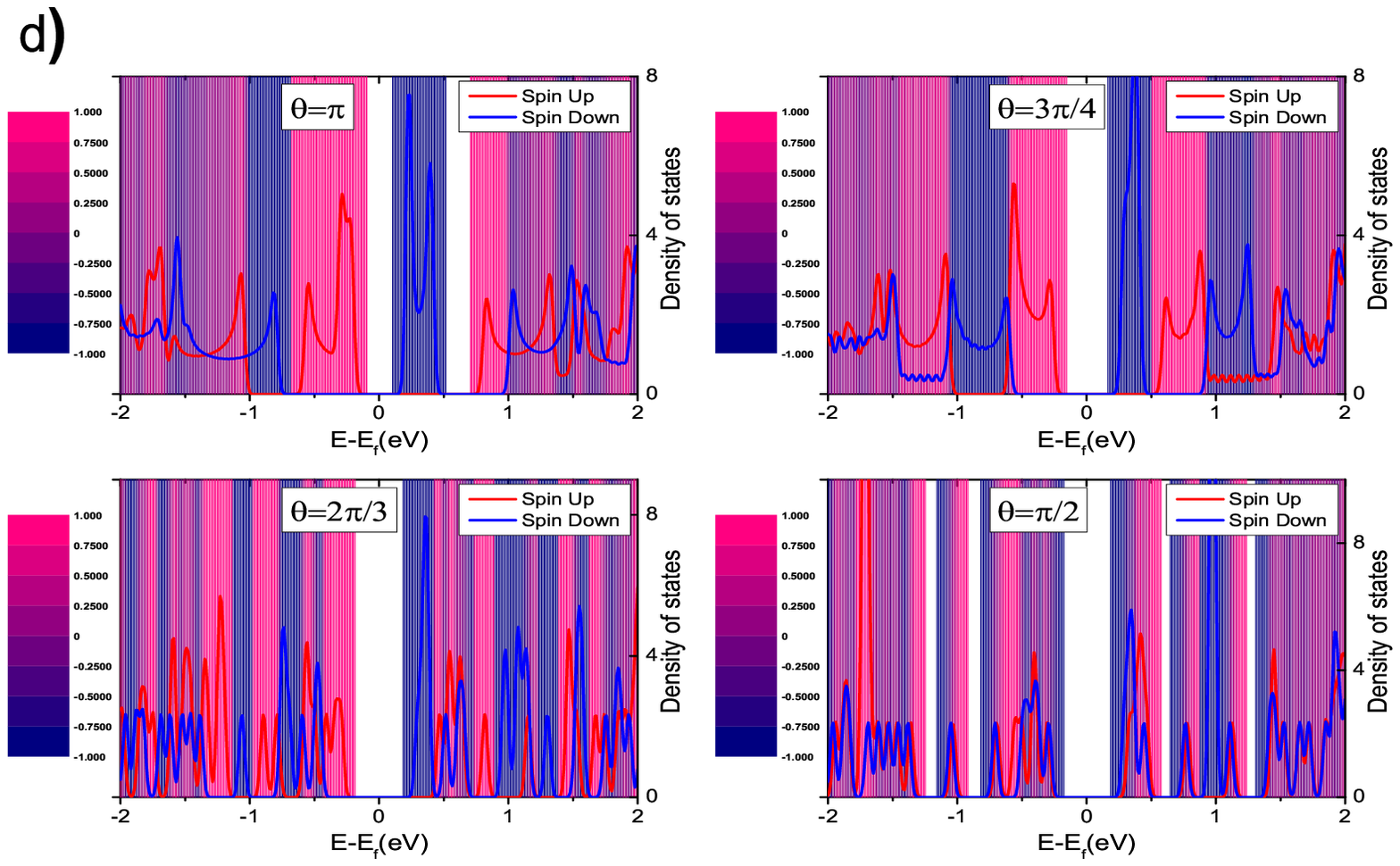}
\caption{(color online) (a)(b) Spin-resolved band structure (left
panel) and density of states (right panel) of the ground state of
TMM and CH$_2$ terminated ZGNR, respectively. (c) Local density of
states(LDOS) of the ground state of CH$_2$ terminated ZGNR. Red for
up spin. (d) Spin-resolved DOS of CH$_2$ terminated ZGNRs with spin
angle $\theta$ equal to $\pi$, $3\pi/4$, $2\pi/3$, and $\pi/2$
respectively. The shaded area presents the spin polarization P(E)
defined in Eq. (\ref{eq:pe}).}

\label{fig3}
\end{figure}

Besides the all-organic ferromagnetic ground state mentioned above,
we find that a large spin-split band gap and the 100\% spin
polarization near the Fermi energy are specific characters of TMM
and CH$_2$ terminated ZGNR. Fig.\ref{fig3}a and b respectively plot
their spin-resolved band structures and density of states (DOS).
From these figures, TMM and CH$_2$ terminated ZGNRs exhibit the
following similar features: 1) both of the spin up and down subbands
manifest semiconductor character with energy gaps $\sim 1
\textrm{eV}$; and 2) the spin splitting between up and down subbands
near the Fermi energy ($E_{\mathrm{F}}$) is about 0.5 eV, which
induces a gapped spin-split DOS with 100\% spin polarization within
a wide energy region near $E_{\mathrm{F}}$. The highest occupied and
lowest unoccupied energy level are proved to be extended states
which can serve a large current in graphene-based transistors. In
Fig.\ref{fig3}c, the local density of states (LDOS, with energy
region $E_{\mathrm{F}}\rightarrow E_{\mathrm{F}}-0.6$ eV) of the
highest (spin up) occupied energy level of CH$_2$ terminated ZGNR is
shown. Towards the practical transistors, those features provide
well-defined conducting ON- and insulating OFF-states, and also the
high spin-polarized current at the ON-state. By applying a positive
or negative gate voltage, one can selectively shift the (purely spin
up) occupied highest energy level or the lowest (purely spin down)
unoccupied one towards the Fermi energy. A 100\% spin polarized
current is thus produced via the gate voltage control.

For comparison, we make some comments on the nonmagnetic side group
terminated ZGNRs in the literature\cite{sodi}. For those materials,
the sizable energy gap and 100\% spin polarization in large energy
range near $E_{\mathrm{F}}$ can not obtained simultaneously. For
example, in the NH$_2$, OH, and COOH terminated ZGNRs, the energy
gap is about 0.3 eV, but the spin splitting of the states near the
$E_{\mathrm{F}}$ is very small. In the NO$_2$ terminated ZGNR,
although the spin splitting is relatively enlarged, the energy gap
reduced to about 0.1 eV, too small to be used in transistors.
Therefore, the present design has an advantage in realizing a large
energy gap and a high spin polarization near the Fermi energy
simultaneously.
%

In spintronics, the spin polarization $P(E)$ may be defined as
\begin{equation}\label{eq:pe}
P(E)=\frac{DOS_{\uparrow}(E)-DOS_{\downarrow}(E)}{DOS_{\uparrow}(E)+DOS_{\downarrow}(E)},
\end{equation}
which can be tuned by changing the spin polarized DOS near
$E_{\mathrm{F}}$. For this purpose, we study the evolution of the
spin polarization near $E_{\mathrm{F}}$ with manually varying the
spin angle $\theta$ in CH$_2$ terminated ZGNR. The result is
demonstrated in Fig. \ref{fig3}d. As mentioned above, the DOS above
and below the $E_{\mathrm{F}}$ are both 100\% spin polarized in a
wide energy range at $\theta=\pi$. However, as $\theta$ decreases
towards $\pi/2$, the high spin polarized region near
$E_{\mathrm{F}}$ shrinks and even almost disappears at
$\theta=\pi/2$. As seen in the figure, the large value of DOS near
$E_{\mathrm{F}}$ during radical rotation can keep the ON-state under
certain gate voltage. Thus, controlling the exchange coupling
strength by radical rotation is an effective way to tune the spin
polarization.

Finally, let us comment on the possibility of taking Lieb's theorem
on the repulsive Hubbard model \cite{Lieb} as the mechanism of the
gapped ferromagnetic state, for the reason that its deduced Klein's
edge \cite{Klein} has been applied to explain the ferromagnetic
state of some other edge terminated GNRs in literatures
\cite{Fuj96,Lakshmi09}. The central point of the Lieb's theorem is
that the ground state of a bipartite lattice and half-filled band
has a net total spin $S$ providing the two sublattice have different
number of sites. By carefully comparing our results with Lieb's
theorem, we conclude that it can not consistently explain all the
features of the ground state here. The main discrepancies are
summarized as follows: 1) it is the strong spin-lattice interaction,
rather than the Hubbard (electron-electron) correlation, stabilizes
the ground state of the present system; 2) the spatial spin order of
the ground state is beyond Lieb's theorem that can only deduce a
nonzero total spin; and 3) our DFT and model calculations have both
omitted the quantum fluctuation of electron-electron interaction
which is the curial part of Hubbard model and Lieb's theorem.
Relative to Lieb's theorem, our mechanism of the antiferromagnetic
exchange together with strong spin-lattice interactions can explain
all of the features of the gapped ferromagnetic state in a
consistent manner, as already demonstrated above.

\emph{Summary}.---In summary, we designed a graphene-based all-organic ferromagnetic semiconductor by terminating ZGNR with organic mangets. A large
spin-split gap with 100\% spin polarized DOS near $E_{\mathrm{F}}$ is
obtained, which is of potential usage in spintronic devices. Combining first-principles calculations with basic model analysis,
we conclude that the mechanism for the gapped ferromagnetic state
is the antiferromagnetic $\pi-\pi$ exchange together with strong spin-lattice
interactions between organic magnet and graphene nanoribbons. By controlling the $\pi-\pi$
interaction strength, the spin polarization of the DOS near $E_{\mathrm{F}}$ can be tuned from 100\% to nearly zero. The fundamental
physics makes it possible to optimally select the organic magnet
towards practical applications.

\begin{acknowledgments}
The authors thank Prof. Shimin Hou of Peking University and Prof. Mingwen Zhao of Shandong University for the insightful discussions and suggestions. Support from National Basic
Research Program of China (Grant No.s 2007CB925001, 2009CB929204 and
2010CB923402) and NSFC of China (Grant No.s 10874100 and 11074303)
are gratefully acknowledged. Part of the computer time is supported
by Physics Laboratory for High Performance Computing, Renmin University
of China. \end{acknowledgments}

\end{document}